\documentclass[twocolumn]{aastex63}

\usepackage[utf8]{inputenc}
\usepackage{amsmath}
\usepackage{amssymb}
\usepackage{graphicx}
\usepackage{epstopdf}
\usepackage{hyperref}
\usepackage{times}

\begin{document}
\title{From the Fermi blazar sequence to the relation between Fermi blazars and $\gamma$-ray Narrow-line Seyfert 1 Galaxies}

\correspondingauthor{Yongyun Chen}
\email{ynkmcyy@yeah,net}

\correspondingauthor{Nan Ding}
\email{orient.dn@foxmail.com}

\author{Yongyun Chen$^{*}$}
\affiliation{College of Physics and Electronic Engineering, Qujing Normal University, Qujing 655011, P.R. China}

\author{Qiusheng Gu}
\affiliation{School of Astronomy and Space Science, Nanjing University, Nanjing 210093, P. R. China}

\author{Junhui Fan}
\affiliation{Center for Astrophysics,Guang zhou University,Guang zhou510006, China}

\author{Hongtao Wang}
\affiliation{School of science, Langfang Normal University, Langfang  065000, P.R. China}

\author{Shaojie Qin}
\affiliation{Middle School of tangtang Town of Xuanwei 655400, P. R. China}

\author{Nan Ding$^{*}$}
\affiliation{School of Physical Science and Technology, Kunming University 650214, P. R. China}

\author{Xiaolin Yu}
\affiliation{School of Astronomy and Space Science, Nanjing University, Nanjing 210093, P. R. China}

\author{Xiaotong Guo}
\affiliation{School of Astronomy and Space Science, Nanjing University, Nanjing 210093, P. R. China}

\author{Dingrong Xiong}
\affiliation{National Astronomical Observatories/Yunnan Observatories, Chinese Academy of Sciences, Kunming 650011,China }

\begin{abstract}
We use the third catalog of blazars detected by Fermi/LAT (3LAC) and $\gamma$-ray Narrow-line Seyfert 1 Galaxies ($\gamma$-NLSy1s) to study the blazar sequence and relationship between them. Our results are as follows: (i) There is a weak anti-correlation between synchrotron peak frequency and peak luminosity for both Fermi blazars and $\gamma$-NLSy1s, which supports the blazar sequence. However, after Doppler correction, the inverse correlation disappeared, which suggests that anti-correlation between synchrotron peak frequency and peak luminosity is affected by the beaming effect. (ii) There is a significant anti-correlation between jet kinetic power and synchrotron peak frequency for both Fermi blazars and $\gamma$-NLSy1s, which suggests that the $\gamma$-NLSy1s could fit well into the original blazar sequence. (iii) According to previous work, the relationship between synchrotron peak frequency and synchrotron curvature can be explained by statistical or stochastic acceleration mechanisms. There are significant correlations between synchrotron peak frequency and synchrotron curvature for whole sample, Fermi blazars and BL Lacs, respectively. The slopes of the correlation are consistent with statistical acceleration. For FSRQs, LBLs, IBLs, HBLs, and $\gamma$-NLS1s, we also find a significant correlation, but in these cases the slopes can not be explained by previous theoretical models. (iv) The slope of relation between synchrotron peak frequency and synchrotron curvature in $\gamma$-NLS1s is large than that of FSRQs and BL Lacs. This result may imply that the cooling dominates over the acceleration process for FSRQs and BL Lacs, while $\gamma$-NLS1s is the opposite.   

\end{abstract}

\section{INTRODUCTION}
Blazars are the most extreme active galactic nuclei (AGN) whose jets are pointing towards us. According to the equivalent width (EW) of the optical emission lines, Blazars are usually divided into two subclasses: flat-spectrum radio quasars (FSRQs) and BL Lac objects (BL Lacs). FSRQs have EW $>$5\AA, while BL Lacs are smaller than this value \citep{Urr95}. Later, some authors used other physical parameters to distinguish FSRQs and BL Lacs. \cite{Ghi11}  used the ratio of broad emission line luminosity to Eddington luminosity to divide the FSRQs and BL Lacs, namely accretion rate. They pointed out that FSRQs have $\rm{L_{BLR}/L_{Edd} \geq 5\times10^{-4}}$, while BL Lacs have $\rm{L_{BLR}/L_{Edd} < 5\times10^{-4}}$. This division between FSRQs and BL Lacs may imply that they have different accretion regime \citep{Sba14}.

\cite{Fos98} proposed the so-called ``blazar sequence": the synchrotron peak luminosity and Compton dominance are anti-correlated with the synchrotron peak frequency. By fitting the spectral energy distributions (SEDs) of blazars, \cite{Ghi98} confirmed the discovery of \cite{Fos98}. They suggested that radiative cooling lead to the formation of a blazar sequence. Some authors supported the blazar sequence \citep{Cav02,Mar03,Mar08,Ghi08,Che11,Fin13,Che14,Xio15}. However, some authors opposed this view \citep{Pad03,Nie06,Pad07,Nie08,Gio12}. The authors mainly considered that the selection effect of the samples lead to the blazar sequence. \cite{Nie08} used a low limit Doppler factor to correct for synchrotron peak frequency and peak  frequency luminosity. They found that the anti-correlation between synchrotron peak frequency and peak frequency luminosity disappeared. \cite{Mey11} used a large sample of radio-loud AGNs to restudy the blazar sequence. They proposed the blazar envelope: Fanaroff–Riley (FR) I radio galaxies and most of BL Lacs are ``weak-jet'' sources, exhibiting radiatively inefficient accretion. However, low synchrotron-peaking (LSP) blazars and FR II radio galaxies are ``strong-jet'' sources, exhibiting radiative efficient accretion. \cite{Fin13} found an anti-correlation exists between Compton dominance and the synchrotron peak frequency by using the Second Large-Area Telescope (2LAT) AGN Catalog \citep{Ack11}. \cite{Mao16} restudied the blazar sequence by using a large sample of blazars from the Roma-BZCAT catalog. They confirmed the original blazar sequence. \cite{Ghi17} used the third Large-Area Telescope AGN Catalog \citep{Ack15} to revisit the blazar sequence. They constructed their average spectral energy distribution (SED) by using $\gamma$-ray luminosity, and found that the synchrotron peak frequency is anti-correlated with $\gamma$-ray luminosity. Their results also support the original blazar sequence.

Besides synchrotron peak frequency and synchrontron peak frequency luminosity, the spectral curvature is also an important physical parameter in the SEDs of blazars. The relationship between the synchrotron peak frequency and the spectral curvature can reflect the acceleration process of particles \citep{Massaro04,Massaro06,Paggi09,Tra09,Tra11}. However, all these previous studies investigated the relationship between
SED peak frequency and its curvature by fitting the SEDs through a log-parabolic function, using observational data with
spectral windows close to the synchrotron peak frequency. The choice to use broad-band fit of the synchrotron emission, without a proper physical model, can introduce a significant bias in the estimate of the curvature. \cite{Che14} predicted the two particle acceleration mechanisms based on the coefficient of relationship between synchrotron peak frequency and spectral curvature. For stochastic acceleration and statistical acceleration, the slope $k$ (1/b$_{sy}$ = $k$log $\nu_{p}$ + c) is 2, 2.5, and 3.33, respectively. They studied this relation by using 43 blazars and found the slope is 2. This result is consistent with stochastic acceleration.       

The $\gamma$-NLSy1s are the mysterious class of the radio-loud AGN. These $\gamma$-NLSy1s show powerful relativistic jets, low black hole mass ($10^{6}-10^{8}M_{\odot}$), and high accretion rate ($0.1-1L_{Edd}$). Some authors thought that their physical properties are similar to blazars. The EW of broad emission line is larger than 5\AA~for all of $\gamma$-NLSy1s \citep{Osh01,Zho07,Yao15b,Rak17}, which may imply that these $\gamma$-NLSy1s can be formally classified as FSRQs. \cite{Fos11} studied the characteristics of the jet of blazars and $\gamma$-NLSy1s. They found that the jet powers of FSRQs and $\gamma$-NLSy1s depend on the black hole mass, while the jet powers of BL Lacs are dependent on the accretion rate. These results suggested that the accretion disk of FSRQs and $\gamma$-NLSy1s are dominated by radiation-pressure, while BL Lacs are dominated by the gas-pressure. \cite{Sun15} found that the jet properties of $\gamma$-NLSy1s resemble that of FSRQs. \cite{Pal18} found that the $\gamma$-NLSy1s and FSRQs occupy the same region in the  Wide-field Infrared Survey Explorer (WISE) color–color diagram. What's more, the $\gamma$-NLSy1s occupy the low black hole mass end of the FSRQs distribution \citep{Pal18}. These $\gamma$-NLSy1s may be the counterpart of powerful FSRQs with low black hole mass \citep{Fos15}. \cite{chen19} studied the relationship between jet power and accretion disk luminosity in blazars and flat-spectrum radio-loud Narrow-line Seyfert 1 galaxies (FRLNLS1s). They found that the slope of such a relation is similar in FRLNLS1s and blazars. According to the SED modeling, \cite{Pal19} found that $\gamma$-NLS1s follow the relation between jet power and accretion luminosity see among blazars. They suggested that the radiation mechanisms of $\gamma$-NLS1s are similar to blazars.

Although many authors have studied the blazar sequence. However, there is no large sample to consider beaming effects when studying the blazar sequence. It has always been controversial about the blazar sequence. At the same time, there is a question: what is the relation between Fermi blazars and $\gamma$-NLS1s? In this paper, we use a large sample of Fermi blazars and $\gamma$-NLSy1s to study blazar sequence and the relation between Fermi blazars and $\gamma$-NLSy1s when considering the beaming effects. The samples are described in Section 2; the results are presented in Section 3; discussions are in Section 4; conclusions are in Section 5. The cosmological parameters $\rm{H_{0}=70kms^{-1}Mpc^{-1}}$, $\rm{\Omega_{m}=0.3}$, and $\rm{\Omega_{\Lambda}=0.7}$ have been adopted in this work.

\begin{figure*}
	\centering
	\includegraphics[width=14cm,height=14cm]{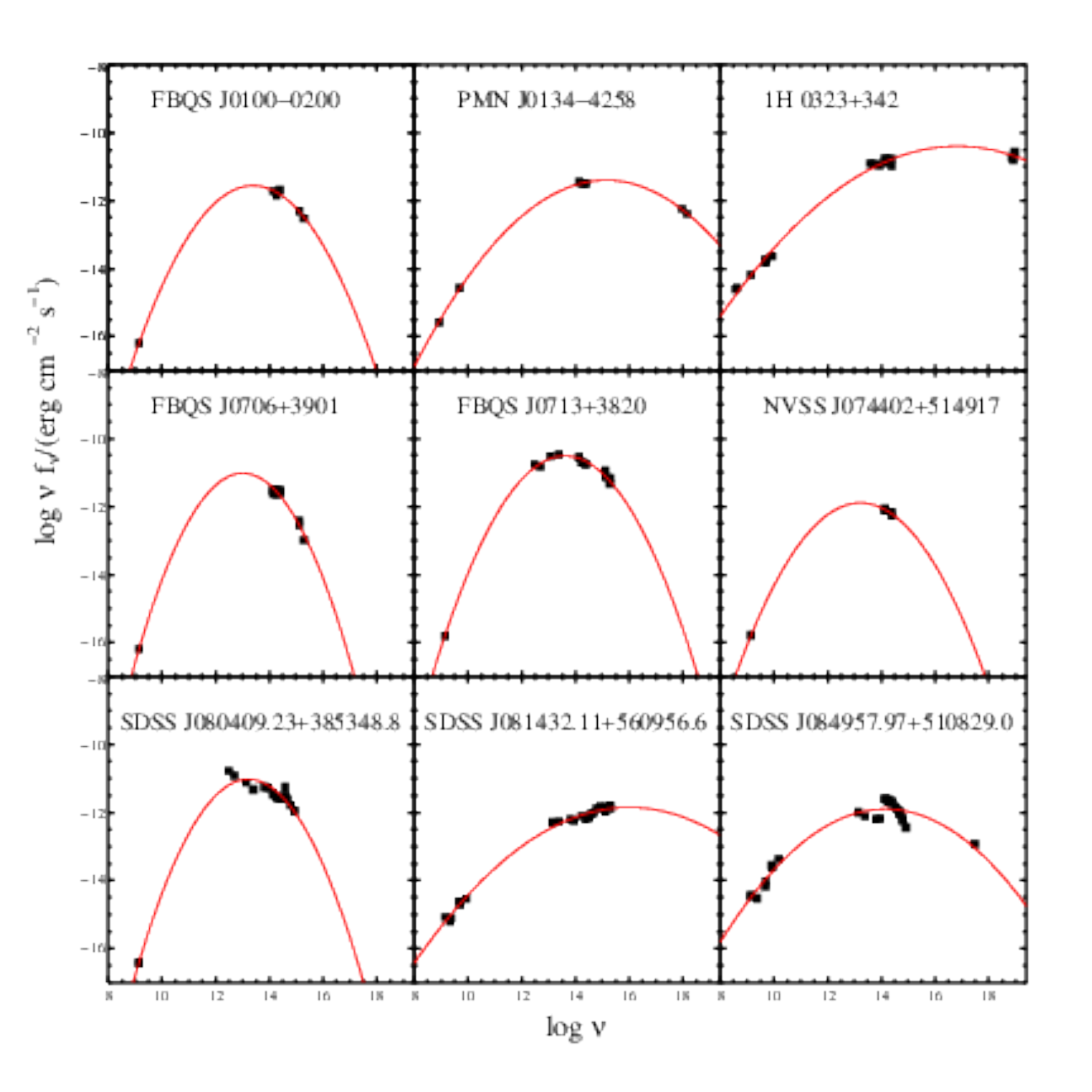}
	\vspace{0pc}
	\caption{The example of SED fitting of $\gamma$-NLS1s.}
	\label{sample-fig1}
\end{figure*}

\section{THE SAMPLE}
\subsection{The sample of Fermi Blazars}
We try to collect a larger sample of Fermi blazars with reliable redshift, synchrotron peak frequency, peak frequency luminosity ($\rm{L_{peak}}$), jet power, and Doppler factor. Firstly,
we consider the sample of \cite{Fan16} to get synchrotron peak frequency and peak frequency luminosity. \cite{Fan16} compiled the multi-wavelength data of 1425 Fermi blazars from 3FGL \citep{Ace15} to calculate their spectral energy distributions (SEDs). They used a parabolic function to fit the multi-wavelength data of 1425 Fermi blazars. The synchrotron peak frequency and peak luminosity were successfully obtained for only 1392 Fermi blazars (461 FSRQs, 620 BL Lacs, and 311 blazars of uncertain type [BCUs]; 999 sources have known redshifts). Secondly, we consider the sample of \cite{Che18} to get jet power and Doppler factor. According to the leptonic model, \cite{Che18} estimated the jet power and Doppler factor of the 1392 Fermi blazars from the catalog of \cite{Fan16}. Finally, we only use the Fermi blazars with reliable redshift and blazars of certain type. Among the 999 sources with known redshift, 75 were of uncertain type. We therefore get 924 Fermi blazars (461 FSRQ, 463 BL Lacs, see Table 1).    

\begin{table*} 
	\begin{center}
		\caption{The physical parameter of Fermi blazars.}	
		\begin{tabular}{llllllllllllllllll}
			\hline
			\hline
			Name  & Class  &  $z$   & $\delta$ &  $\log P_{jet}$ & $\log \nu_{peak}$ & $\log L_{peak}$ & $b$  \\
			(1) & (2) & (3) & (4) & (5) & (6) & (7) & (8)\\
			\hline			
J0001.4+2120 	&	F 	&	1.106 	&	10.7	&	45.9	&	16.79 	&	45.70 	&	0.05 	\\
J0004.7-4740 	&	F 	&	0.880 	&	12.3	&	46.2	&	14.14 	&	46.20 	&	0.12 	\\
J0006.4+3825 	&	F 	&	0.229 	&	5.6	&	45.4	&	14.03 	&	44.65 	&	0.11 	\\
J0008.0+4713 	&	IB	&	0.280 	&	18.4	&	46	&	14.52 	&	44.46 	&	0.12 	\\
J0008.6-2340 	&	IB	&	0.147 	&	51.1	&	45.8	&	15.09 	&	44.01 	&	0.10 	\\
J0013.9-1853 	&	IB	&	0.095 	&	29	&	44.8	&	14.96 	&	44.37 	&	0.13 	\\
J0016.3-0013 	&	F 	&	1.577 	&	6.7	&	46.4	&	13.58 	&	45.58 	&	0.09 	\\
J0017.6-0512 	&	F 	&	0.227 	&	5	&	45.1	&	14.48 	&	44.63 	&	0.11 	\\
J0018.4+2947 	&	HB	&	0.100 	&	14.3	&	44.9	&	16.60 	&	43.44 	&	0.06 	\\
J0023.5+4454 	&	F 	&	1.062 	&	7.6	&	46.5	&	12.78 	&	44.73 	&	0.13 	\\
J0024.4+0350 	&	F 	&	0.545 	&	25.5	&	45.6	&	13.09 	&	45.37 	&	0.26 	\\
J0030.3-4223 	&	F 	&	0.495 	&	6.6	&	45.8	&	14.14 	&	45.43 	&	0.12 	\\
J0032.3-2852 	&	IB	&	0.324 	&	71.9	&	46.5	&	13.97 	&	44.92 	&	0.15 	\\
J0033.6-1921 	&	HB	&	0.610 	&	12.3	&	45.1	&	15.74 	&	45.96 	&	0.11 	\\
J0035.2+1513 	&	IB	&	0.250 	&	27.5	&	45.5	&	15.04 	&	44.77 	&	0.12 	\\
J0035.9+5949 	&	HB	&	0.086 	&	14.3	&	45.6	&	18.46 	&	44.21 	&	0.04 	\\
J0037.9+1239 	&	IB	&	0.089 	&	25.8	&	45.6	&	14.24 	&	43.92 	&	0.14 	\\
J0038.0+0012 	&	LB	&	0.740 	&	14.3	&	46.4	&	12.89 	&	45.70 	&	0.25 	\\
J0038.0-2501 	&	F 	&	0.498 	&	15.1	&	45.9	&	13.26 	&	45.68 	&	0.19 	\\

\hline
\end{tabular}
\tablecomments{Column (1) gives the 3FGL Name. Column (2) is the class of sources. Column (3) gives the redshift. Column (4) is the Doppler factor. Column (5) is the jet power in units of $\rm{erg~s^{-1}}$. Column (6) is the synchrotron peak frequency. Column (7) gives the peak frequency luminosity in units of $\rm{erg~s^{-1}}$. Column (8) is the curvature b. The F is FSRQs; IB is intermediate synchrotron peaked BL Lacs; HB is high synchrotron peaked BL Lacs; LB is low synchrotron peaked BL Lacs. This table is published in its entirety in the electronic edition. A portion is shown here for guidance. The data is download in http://cdsportal.u-strasbg.fr/my-data/.} 
\end{center}
\label{para}
\end{table*}

\subsection{The $\gamma$-ray narrow-line seyfert 1 galaxies}
We try to collect a large sample of $\gamma$-NLS1s with reliable redshift, jet power, and Doppler factor. We consider the sample of \cite{Pal19} to get jet power and Doppler factor. \cite{Pal19} compiled the largest sample of $\gamma$-NLS1s to study their physical properties. They got the jet power and Doppler factor of 16 $\gamma$-NLS1s based on the leptonic model. Following the work of \cite{Fan16}, we use a parabolic function to fit the quasi-simultaneous multi-wavelength data of 16 $\gamma$-NLS1s and get their synchrotron peak frequency and peak frequency luminosity. The fitting formula is as follows

\begin{equation}
\rm{log(\nu F_{\nu}) = -b(log\nu-log\nu_{peak})^{2} + log\nu_{peak} f_{\nu_{p}}},
\end{equation}
where $b$ is the spectral curvature, $\rm{log\nu_{peak}}$ is the peak frequency and $\rm{log(\nu_{peak}f_{\nu_{peak}})}$ is the peak flux. The sample of $\gamma$-NLS1s is shown in Table 2. Figure 1 shows the example of SED of $\gamma$-NLS1s. 

\begin{table*} 
	\begin{center}
		\caption{The physical parameter of 16 $\gamma$-NLS1s.}	
		\begin{tabular}{llllllllllllllllll}
			\hline
			\hline
			Name  &   $z$   & $\delta$ &  $\log P_{jet}$ & $\log \nu_{peak}$ & $\log L_{peak}$ & $b$  \\
            (1) & (2) & (3) & (4) & (5) & (6) & (7) \\
            \hline			
1H 0323+342	&	0.061	&	13.6	&	45.82	&	14.98	&	45.91	&	0.99	\\
SBS 0846+513 	&	0.584	&	19.1	&	46.05	&	13.65	&	45.15	&	0.88	\\
CGRaBS J0932+5306 	&	0.597	&	14.7	&	46.54	&	15.22	&	44.93	&	1.25	\\
GB6 J0937+5008 	&	0.275	&	15.4	&	46.41	&	15.04	&	44.38	&	1.14	\\
PMN J0948+0022 	&	0.585	&	15.7	&	47.11	&	15.3	&	45.00	&	1.19	\\
TXS 0955+326 	&	0.531	&	12.3	&	46.68	&	14.61	&	45.79	&	1.63	\\
FBQS J1102+2239 	&	0.453	&	19	&	45.86	&	13.31	&	44.98	&	0.57	\\
CGRaBS J1222+0413 	&	0.966	&	16.5	&	47.59	&	14.95	&	45.89	&	1.2	\\
SDSS J124634.65+023809.0 	&	0.362	&	17.8	&	45.67	&	15.67	&	44.52	&	1.11	\\
TXS 1419+391	&	0.49	&	13.6	&	46.77	&	15.15	&	44.73	&	1.13	\\
PKS 1502+036	&	0.407	&	17.2	&	46.08	&	13.34	&	44.79	&	0.89	\\
TXS 1518+423	&	0.484	&	17.8	&	46.32	&	15.9	&	44.56	&	1.62	\\
RGB J1644+263	&	0.145	&	14.7	&	45.91	&	14.56	&	43.80	&	0.96	\\
PKS 2004-447	&	0.24	&	17.2	&	45.91	&	14.67	&	43.77	&	1.27	\\
TXS 2116-077	&	0.26	&	17.2	&	45.92	&	14.46	&	43.90	&	1.11	\\
PMN J2118+0013 	&	0.463	&	14.7	&	45.99	&	16.55	&	44.62	&	1.35	\\
\hline
\end{tabular}
\tablecomments{ Column (1) gives the Name. Column (2) gives the redshift. Column (3) is the Doppler factor. Column (4) is the jet power in units of $\rm{erg~s^{-1}}$. Column (5) is the synchrotron peak frequency. Column (6) gives the peak frequency luminosity in units of $\rm{erg~s^{-1}}$. Column (7) is the curvature b.} 
\end{center}
\label{para}
\end{table*}

\section{Results}
\subsection{The correlation between synchrotron peak frequency and peak frequency luminosity}
We study the correlation between synchrotron peak frequency and synchrotron peak frequency luminosity using redshift-corrected values. The synchrotron peak frequency luminosity is estimated by using the following formula

\begin{equation}
\rm{L_{peak} = 4\pi d_{L}^{2}\nu_{peak} F_{\nu,peak}}
\end{equation}
where $d_{L}$ is luminosity distance, $d_{L}(z) = \frac{c}{H_{0}}(1+z)\int_{0}^{z}[\Omega_{\rm{\Lambda}}+\Omega_{\rm{m}}(1+z')^{3}]^{-1/2}dz'$ \citep{Ven09}. The redshift-corrected synchrotron peak frequency is calaulated by using formula 
 
\begin{equation}
\rm{\nu_{peak}^{sy} = \nu_{peak} (1+z)}
\end{equation}
 
 \begin{figure}
\centering
\includegraphics[width=8cm,height=8cm]{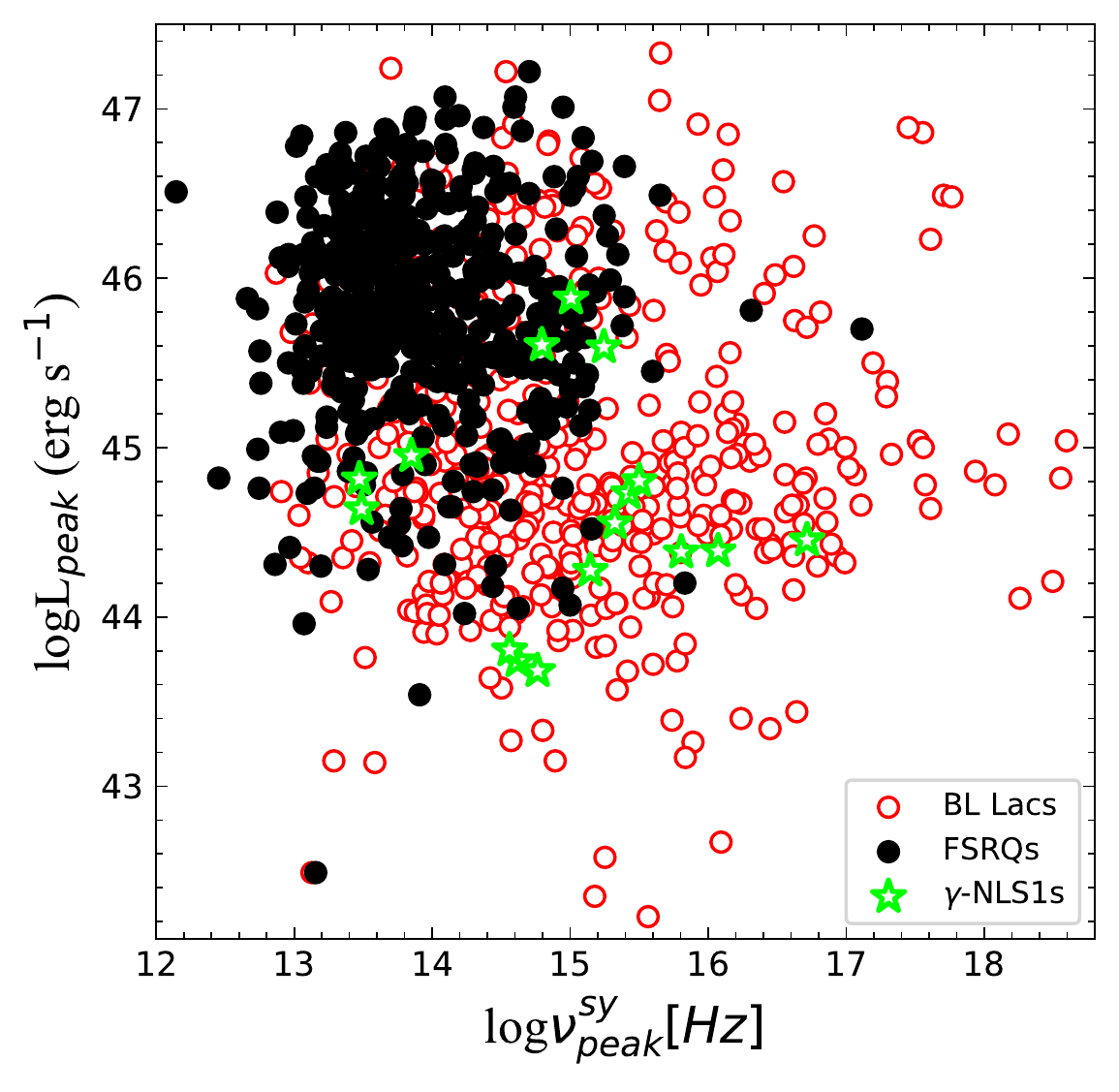}
\vspace{0pc}
\caption{The Synchrotron peak luminosity versus peak frequency for whole sample. The black filled circle is FSRQs.The red empty circle is BL Lacs. The green star is $\gamma$-NLS1s.}
\label{sample-fig2}
\end{figure}

Figure 2 shows the relationship between the synchrotron peak frequency luminosity and synchrotron peak frequency. We do not find a ``L" or ``V" shape in this figure. Fermi blazars and $\gamma$-NLS1s are located in the same region. The $\gamma$-NLS1s tend to have lower synchrotron peak frequency luminosity than FSRQs. The results of Pearson analysis show that there is a weak negative correlation between $L_{peak}$ and $\nu_{peak}^{sy}$ for the whole sample (N = 940, r = -0.25, P = 1.58$\times10^{-14}$). The scatter of this correlation is $\sigma$= 0.86 dex. The Analysis of Variance (ANOVA) is used to test the results of linear regression, which shows that it is valid for the results of linear regression(value F =60.92, probability P = 1.58$\times10^{-14}$). At the same time, we also use Kendall and Spearman tests to analyze these correlations besides Pearson. The results of Kendall (r =-0.19, P = 4.81$\times10^{-19}$) and Spearman (r =-0.29, P = 2.89$\times10^{-20}$) tests show that there is also a weak anti-correlation between  $L_{peak}$ and $\nu_{peak}^{sy}$ for whole sample. 

\begin{figure}
\centering
\includegraphics[width=8cm,height=15cm]{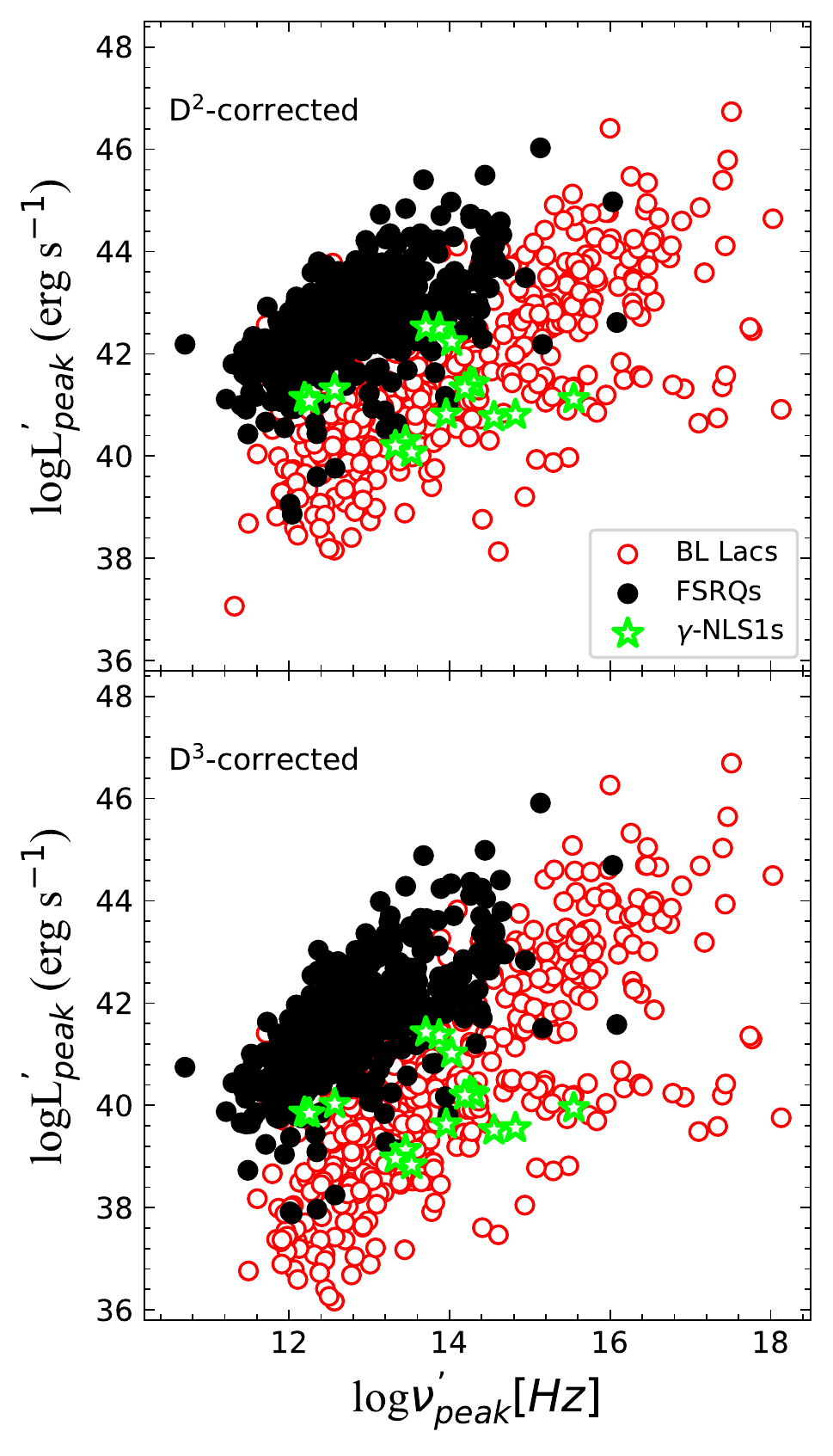}
\vspace{0pc}
\caption{Doppler-corrected synchrotron peak luminosity versus Doppler-corrected synchrotron peak frequency for Fermi blazars. In the top panel, the datapoints are $D^{2}$-corrected. In the bottom panel, the datapoints are D$^{3}$-corrected.}
\label{sample-fig3}
\end{figure}

\cite{Nie08} proposed that the anti-correlation between the synchrotron peak frequency luminosity and synchrotron peak frequency is affected by the beaming effect.  Therefore, we use Doppler factor to correct for synchrotron peak frequency and synchrotron peak frequency luminosity. The Doppler-corrected synchrotron peak frequency is performed using equation
\begin{equation}
\nu_{peak}^{'} = \frac{\nu_{peak}^{sy}}{\delta}
\end{equation}
where $\nu_{peak}^{'}$ indicates the $\delta$-corrected $\nu_{peak}$ in the rest frame. The Doppler-corrected synchrotron peak frequency luminosities are performed using the following formula

\begin{equation}
 L_{peak}^{'} =  \frac{L_{peak}}{{\delta}^{P}}
\end{equation}
 where P= 2+ $\alpha$ is a continuous jet and P= 3+ $\alpha$ is a spherical jet \citep{Urr95}, spectral index $\alpha$=1.

According to equation (4) and (5), the Doppler-corrected synchrotron peak frequency luminosity and peak frequency can be obtained ($D^{2}$-correction and $D^{3}$-correction indicates P=2+$\alpha$ and P=3+$\alpha$, respectively ). The Doppler-corrected synchrotron peak frequency luminosity versus the Doppler-corrected synchrotron peak frequency is shown in Figure 3. The top panel of Figure 3 (P=2+$\alpha$) shows that there are significant positive correlations for whole sample (r = 0.39, P = 8.77$\times10^{-35}$). The Analysis of Variance (ANOVA) is used to test the results of linear regression, which shows that it is valid for the results of linear regression(value F = 164.3, probability P = 8.77$\times10^{-35}$). The results of Kendall (r =0.25, P = 2.42$\times10^{-30}$) and Spearman (r =0.36, P = 7.20$\times10^{-30}$) tests show that there is also a correlation between them for whole sample. We can find that there are also significant positive correlations between Doppler-corrected synchrotron peak luminosity and the Doppler-corrected synchrotron peak frequency from the bottom of Figure 3 (P=3+$\alpha$) for whole sample (r = 0.46, P = 1.83$\times10^{-50}$). The Analysis of Variance (ANOVA) is used to test the results of linear regression, which shows that it is valid for the results of linear regression(value F = 252.1, probability P = 1.83$\times10^{-50}$). The results of Kendall (r =0.30, P = 4.68$\times10^{-43}$) and Spearman (r =0.42, P = 5.68$\times10^{-42}$) tests show that there is also a correlation between them for whole sample.

At the same time, we should also pay attention to the so-called ``bulk Lorentz factor crisis'' in particular regarding the HBLs and TeV detected HBLs when Doppler correction. The one-zone synchrotron self-Compton process (SSC) model requires much higher Lorentz/Doppler factor values \citep{Tavecchio98, Konopelko03, Sauge04, Krawczynski01}. However, the TeV blazars have no clear superluminal motion, which implies a low Lorentz/Doppler factor in Tev blazars \citep{Piner04,Henri06}. In our sample, 43 of the 924 sources are Tev blazars. The 43 TeV blazars include 4 FSRQs, 8  intermediate synchrotron peaked BL Lacs (IBLs) and 31 high synchrotron peaked BL Lacs (HBLs). We find that the percentage of TeV blazars is relatively low in our sample, only 4.65\%. Therefore, the so-called Lorentz factor crisis does not have a significant impact on our main results when we perform Doppler correction.  

\subsection{Jet power versus synchrotron peak frequency}
The relation between jet kinetic power and the synchrotron peak frequency is shown in Figure 4. From a Pearson analysis, we find that there is a significant anti-correlation between jet kinetic power and the synchrotron peak frequency for the whole sample (r = -0.57, P = 2.75$\times10^{-81}$). The Analysis of Variance (ANOVA) is used to test the results of linear regression, which shows that it is valid for the results of linear regression(value F = 445.9, probability P = 2.75$\times10^{-81}$). The results of Kendall (r =-0.38, P = 2.13$\times10^{-65}$) and Spearman (r =-0.54, P = 1.04$\times10^{-71}$) tests show that there is also a significant anti-correlation between them for whole sample. What's more, the $\gamma$-NLS1s follow the blazar sequence.

\begin{figure}
\centering
\includegraphics[width=8cm,height=8cm]{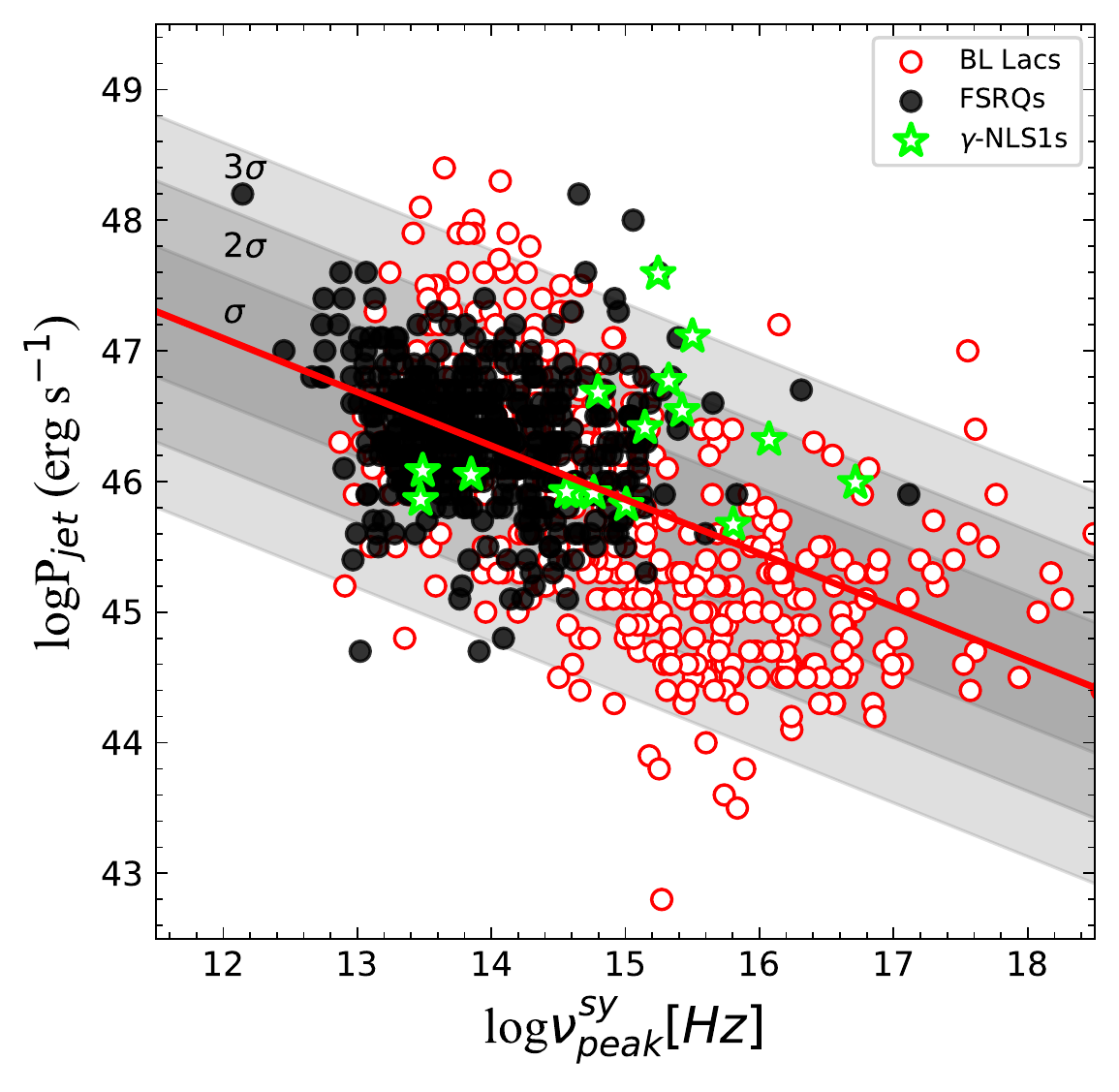}
\vspace{0pc}
\caption{Jet power versus synchrotron peak frequency for whole sample. Shaded areas correspond to 1$\sigma$, 2$\sigma$ and 3$\sigma$ (vertical) dispersion, where $\sigma$ = 0.50 dex. The red line is the least-squares best fit ($\rm{logP_{jet} = -0.41 log\nu_{peak}^{'} + 52.03}$). The meanings of the different symbols are the same as in Figure 2.}
\label{sample-fig4}
\end{figure}

\begin{figure}
\centering
\includegraphics[width=8cm,height=8cm]{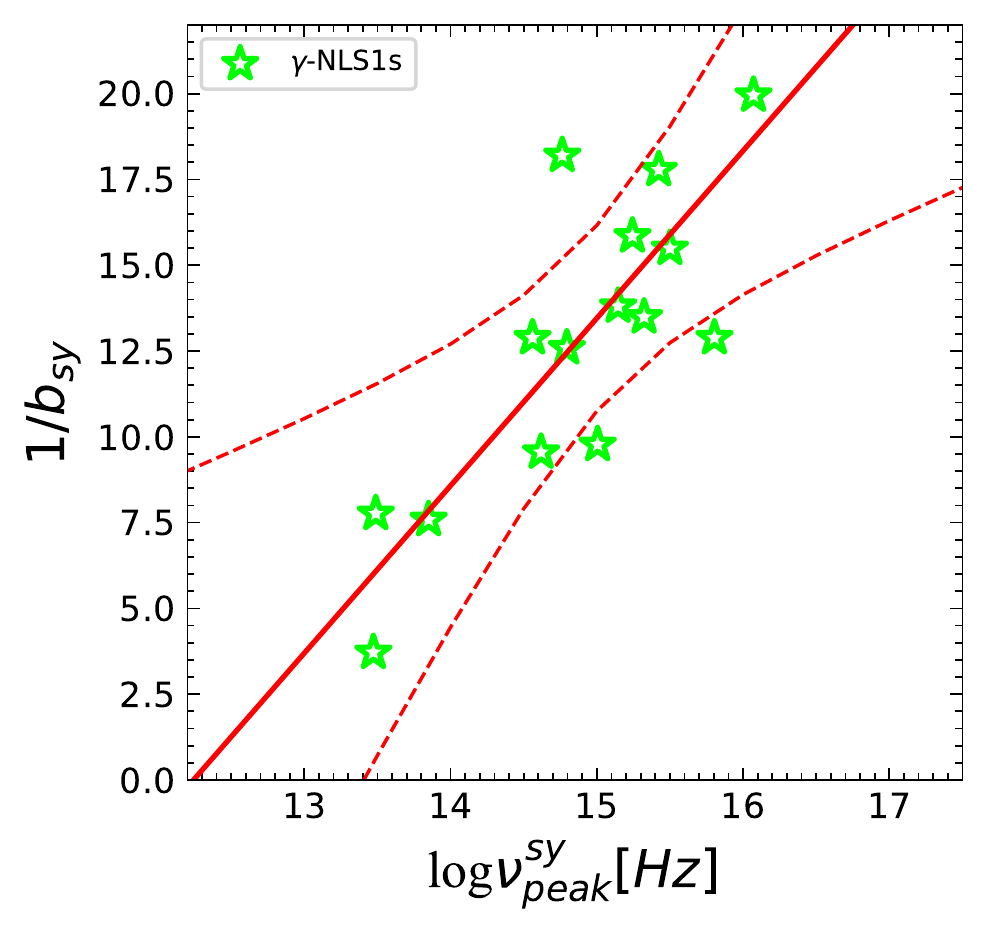}
\vspace{0pc}
\caption{The synchrotron peak frequency versus synchrotron curvature for $\gamma$-NLS1s. The red line is the least-squares best fit ($\rm{1/b_{sy} = (4.87\pm0.76)log\nu_{peak}^{sy} + (-59.79\pm11.45)}$). The red dashed line is the 3$\sigma$ confidence level. The meanings of the different symbols are the same as in Figure 2.}
\label{sample-fig5}
\end{figure}

\subsection{Synchrotron peak frequency versus synchrotron curvature
}
Figure 5 shows the relationship between the synchrotron peak frequency and  synchrotron curvature for $\gamma$-NLS1s. Here we use $1/b_{sy}$ to represent the synchrotron curvature because it will be convenient to compare with the theoretical results (see \cite{Che14}). From a Pearson correlation analysis, there is a significant correlation between synchrotron peak frequency and synchrotron curvature for $\gamma$-NLS1s (r = 0.86, p = 1.67$\times10^{-5}$). The Analysis of Variance (ANOVA) is used to test the results of linear regression, which shows that it is valid for the results of linear regression (value F = 40.88, probability P = 1.67$\times10^{-5}$). The results of Kendall (r =0.62, P = 0.0007) and Spearman (r =0.78, P = 0.0003 ) tests show that there is also a significant correlation between them for whole sample.

\section{Discussions}
\subsection{The Fermi blazar sequence}
In this paper, we use a large sample of Fermi blazars to study the beaming effects on the blazar sequence. \cite{Nie06} and \cite{Mey11} found a ``V" or ``L" shape in the diagrams of $\rm{L_{peak}}$ versus $\rm{\nu_{peak}}$. However, we do not see this shape in Figure 2. \cite{Fin13} studied blazar sequence by using the 352 second LAT sample. They used the empirical relations of \cite{Abd10} to estimate the $\rm{\nu_{peak}}$ and $\rm{L_{peak}}$ and found a significant anti-correlation between the $\rm{\nu_{peak}}$ and $\rm{L_{peak}}$. By comparing Figure 2 with Figure 2 of \cite{Fin13}, we found that Figure 2 is a bit different from the work of \cite{Fin13}. Our results has a larger dispersion than them. These may be due to the different methods. Our sample is larger than that of them. Moreover, we find a weak anti-correlation between synchrotron peak frequency luminosity and the synchrotron peak frequency for the whole sample. These results support the blazar sequence.    

The blazar sequence may be affected by the Doppler factor \citep{Nie08}. In the work of \cite{Nie08}, the Doppler factor is estimated by using the variability of radio flux, corresponding a lower limit to the Doppler factor. We use the Doppler factor derived from the synchrotron self-Compton (SSC) emission model \citep{Che18}. From Figure 3, we find that after being Doppler-corrected, for all Fermi blazars, the correlations between $\rm{L_{peak}}$ and $\nu_{peak}^{sy}$ become significant positive correlations, i.e. the anti-correlation between $\rm{L_{peak}}$ and $\nu_{peak}^{sy}$ disappears, which is consistent with the result of \cite{Nie08}. The observational anti-correlation between $\rm{L_{peak}}$ and $\nu_{peak}^{sy}$ is affected by Doppler beaming factor.   

Because synchrotron peak luminosity is strongly affected by the beaming effect. Thus, we study the relationship between intrinsic jet power (jet kinetic power) and the synchrotron peak frequency. We find that there is a significant anti-correlation between $\rm{P_{jet}}$ and $\rm{\nu_{peak}^{sy}}$ for Fermi blazars and $\gamma$-NLS1s, which supports the blazar sequence, i.e. stronger radiative cooling for higher jet power sources results in smaller energies of the electrons emitting at the peaks.

\subsection{The relation between Fermi blazars and $\gamma$-NLS1s}
\cite{Pal13} found that the physical properties of $\gamma$-NLS1s PKS 1502+036 and PKS 2004-447 located between FSRQs and BL Lacs, which imply that theses two sources may belong to the blazar sequence. \cite{Fos17} thought that the blazar evolutionary sequence should include NLSy1s. They proposed the evolutionary sequence from NLS1s to BL Lacs, NLS1s$\rightarrow$FSRQs$\rightarrow$BL Lacs, namely from small-mass highly-accretion to large-mass low accreting black hole. We thus study the relation between Fermi blazars and $\gamma$-NLS1s. We fit the SEDs of $\gamma$-NLS1s to get the synchrotron peak frequency and peak frequency luminosity. There is an anti-correlation between the synchrotron peak frequency and peak frequency luminosity for both Fermi blazars and $\gamma$-NLS1s. The $\gamma$-NLS1s follow the synchrotron peak frequency and peak frequency luminosity relation seen among Fermi blazars (Figure 2). At the same time, we also consider the relationship between jet kinetic power and synchrotron peak frequency (Figure 4). There is a significant anti-correlation between jet kinetic power and synchrotron peak frequency for both Fermi blazars and $\gamma$-NLS1s. The $\gamma$-NLS1s follow the jet kinetic power and synchrotron peak frequency relation seen among Fermi blazars. Our results suggest that these $\gamma$-NLS1s could fit well into the traditional blazar sequence. \cite{Ghi08} proposed that the jet power and the SED of blazars are closely related to the two mian physical parameters of accretion process, namely black hole mass and accretion rate. The radiative cooling leads to the observational phenomenon of blazar sequence \citep{Ghi98}. The FSRQs have high accretion rate, which leads to the fast cooling of relativistic electrons. FSRQs have low synchronous peak frequency and high jet power. However, BL Lacs have low accretion rate, which leads to the slow cooling of relativistic electrons. BL Lacs have  high synchronous peak frequency and lower jet power. Some works have found that $\gamma$-NLS1s have high accretion rate \citep{Fos17,chen19}. The accretion rate of $\gamma$-NLS1s is similar to FSRQs, which imply that relativistic electrons of the jet of FSRQs and $\gamma$-NLS1s are fast cooling.  The $\gamma$-NLS1s may belong to the Fermi blazar sequence.  

\subsection{Particle acceleration mechanisms for Fermi blazars and  $\gamma$-NLS1s}
The correlation between log$\nu_{peak}^{sy}$ and $1/b_{sy}$ can be explained by two different scenarios, namely the statistical acceleration and the stochastical acceleration mechanisms \citep{Che14}. Following the same approach of \cite{Che14}, i.e. to investigate the particle acceleration. We investigate the relationship between  log$\nu_{peak}^{sy}$ and $1/b_{sy}$ for Fermi blazars and $\gamma$-NLS1s. The results are shown in Table 3. We get the relationship between log$\nu_{peak}^{sy}$ and $1/b_{sy}$ for whole sample (N = 940, $r$=0.79, F = 1566, $P=3.16\times10^{-202}$)
\begin{equation}
\rm{1/b_{sy} = (2.44\pm0.06)log\nu_{peak}^{sy} + (-26.13\pm0.89)}
\end{equation}  
and for Fermi blazars (N = 924, $r$=0.79, F =1580, $P=4.27\times10^{-202}$), 
\begin{equation}
\rm{1/b_{sy} = (2.40\pm0.06)log\nu_{peak}^{sy} + (-25.53\pm0.88)}
\end{equation}  
and for FSRQs (N = 461, $r$=0.80, F = 833.5, $P=2.98\times10^{-105}$),
\begin{equation}
\rm{1/b_{sy} = (3.69\pm0.13)log\nu_{peak}^{sy} + (-42.89\pm1.79)}
\end{equation}
and for BL Lacs (N = 463, $r$=0.85, F = 1229, $P=3.84\times10^{-132}$), 
\begin{equation}
\rm{1/b_{sy} = (2.56\pm0.07)log\nu_{peak}^{sy} + (-28.75\pm1.11)}
\end{equation}
and for low synchrotron peaked BL Lacs (LBLs, N = 83, $r$=0.51, F = 28.96, $P=7.02\times10^{-7}$), 
\begin{equation}
\rm{1/b_{sy} = (3.46\pm0.64)log\nu_{peak}^{sy} + (-40.55\pm8.73)}
\end{equation}
and for intermediate synchrotron peaked BL Lacs (IBLs, N = 214, F = 26.13, $r$=0.33, $P=7.12\times10^{-7}$),
\begin{equation}
\rm{1/b_{sy} = (1.67\pm0.33)log\nu_{peak}^{sy} + (-15.72\pm4.79)}
\end{equation} 
and for high synchrotron peaked BL Lacs (HBLs, N = 166, $r$=0.80, F = 305.9, $P=2.48\times10^{-39}$),
\begin{equation}
\rm{1/b_{sy} = (3.09\pm0.18)log\nu_{peak}^{sy} + (-37.33\pm2.89)}
\end{equation}
and for $\gamma$-NLS1s ($r$=0.86, F = 40.88, $P=1.67\times10^{-5}$),
\begin{equation}
\rm{1/b_{sy} = (4.87\pm0.76)log\nu_{peak}^{sy} + (-59.79\pm11.45)}
\end{equation} 

\begin{table*} 
	\begin{center}
		\caption{The Results of Pearson correlation analysis for sample.}	
		\begin{tabular}{llllllllllllllllll}
			\hline
			\hline
			sample  &      &  $\rm{1/b_{sy} = A\log\nu_{peak}^{sy} + B}$ &   &  \\
			\cline{2-5}
			&      A    &   B   &     r    &    p    &   F \\
			\hline
			Whole sample  & 2.44$\pm$0.06 &  -26.13$\pm$0.89 &   0.79 & $3.16\times10^{-202}$ & 1566 \\
			Fermi blazars  & 2.40$\pm$0.06 & -25.53$\pm$0.88  &0.79 & $4.27\times10^{-202}$ & 1580 \\
			FSRQs  & 3.69$\pm$0.13 &  -42.89$\pm$1.79  &0.80 & $2.98\times10^{-105}$ & 833.5 \\
			BL Lacs  & 2.56$\pm$0.07 &  -28.75$\pm$1.11    & 0.85 & $3.84\times10^{-132}$ & 1229 \\
			LBLs  & 3.46$\pm$0.64 & -40.55$\pm$8.73     &  0.51 & $7.02\times10^{-7}$ & 28.96 \\
			IBLs  & 1.67$\pm$0.33 &  -15.72$\pm$4.79  &0.33 & $7.12\times10^{-7}$ & 26.13 \\
			HBLs  & 3.09$\pm$0.18 &  -37.33$\pm$2.89    &  0.80 & $2.48\times10^{-39}$ & 305.9 \\
			$\gamma$-NLS1s  & 4.87$\pm$0.76 &  -59.79$\pm$11.45   &0.86 & $1.67\times10^{-5}$ & 40.88 \\

			\hline
		\end{tabular}
		\tablecomments{The A is slope; B is the intercept; r is correlation coefficient; p is significance level (p$<$0.01); F is statistical testing of linear regression.}
	\end{center}
	\label{para}
\end{table*}

We find that the slopes of the correlation between synchrotron peak frequency and synchrotron curvature of whole sample ($k_{whole~sample}=2.44\pm0.06$), Fermi blazars ($k_{Fermi~blazars}=2.40\pm0.06$) and BL Lacs ($k_{BL Lacs}=2.56\pm0.07$) are consistent with statistical acceleration for the case of energy-dependent acceleration probability. However, for FSRQs, LBLs, IBLs, HBLs, and $\gamma$-NLS1s, the slopes of the correlation are not consistent with any theoretical values ($k$=5/2, 10/3 and 2).        

\cite{Che14} used a sample of 43 blazars to study the correlation between synchrotron peak frequency and synchrotron curvature. They got that the slope of the correlation was $2.04\pm0.03$, which is consistent with the stochastic acceleration. The sample of \cite{Che14} was too small to separate them into FSRQs, BL Lacs, LBLs, IBLs, and HBLs. Maybe that's why they didn't find different slopes between FSRQs, BL Lacs, LBLs, IBLs, HBLs. At the same time, our results are different from their results, which may be due to the difference in the number of samples.    

The slope of the correlation for FSRQs $k_{FSRQs} = 3.69\pm0.13$ is close to 10/3, which can be explained by statistical particle acceleration for the case of fluctuation of the fractional
acceleration gain. \cite{Xue16} studied the relation between synchrotron peak frequency and synchrotron curvature by using a sample of the second LAT AGN catalogue (2LAC). They found that the slope of the correlation for FSRQs is $k_{FSRQs} = 3.69\pm0.24$. Our results are consistent with theirs. We find that the slope of the correlation for IBLs $k_{IBLs} = 1.67\pm0.33$ is close to 2, which can also be explained by stochastic particle acceleration. \cite{Kap20} studied the X-ray spectral of BL Lacs and found the stochastic acceleration in the relativistic jets of BL Lacs. The slopes of the correlation for LBLs  $k_{LBLs} = 3.46\pm0.64$ and HBLs $k_{HBLs} = 3.09\pm0.18$ are close to 10/3, which can be explained by statistical particle acceleration for the case of fluctuations of the fractional acceleration gain. 

The slope of the correlation for $\gamma$-NLS1s $k_{\gamma-NLS1s} = 4.87\pm0.76$ is not close to any theoretical values. Its slope is slightly large than that of FSRQs and BL Lacs. These results may be explained in the framework of acceleration and cooling processes \citep{Tra11}. In the acceleration process, there is a significant dispersion on the energy gain, leading to a momentum diffusion term, a decreasing curvature (namely increasing of $1/b_{sy}$) leads to a shift of the peak frequency toward higher peak frequency. Hence, the correlation
between the peak frequency and curvature is negative. However, the slope of this correlation can change when the cooling dominates over the acceleration process \citep{Tra11,Kal19}. \cite{Tra11} suggested that the magnetic field plays an important role in the evolution of the spectral parameters. They proposed that when the magnetic field is weak, the evolution of the particles around the peak is dominated by the
acceleration process, while it is driven by cooling for strong magnetic field. \cite{Pal19} found that the average magnetic field derived for $\gamma$-NLS1s is relatively lower ($\langle B \rangle = 0.91\pm0.33$ G) compared to Fermi blazars (1.83$\pm$0.25 G). These results may imply that the cooling dominates over the acceleration process for Fermi blazars, while the acceleration dominates over the cooling process for $\gamma$-NLS1s. The slopes of $\gamma$-NLS1s, FSRQs and BL Lacs seems to form an evolutionary sequence, $\gamma$-NLS1s$\rightarrow$FSRQs$\rightarrow$BL Lacs, namely from acceleration (high slope) to cooling process (low slope). \cite{Fos17} thought that the evolutionary sequence $\gamma$-NLS1s$\rightarrow$FSRQs$\rightarrow$BL Lacs may be the different stages of the cosmological evolution of the same type of source (young$\rightarrow$adult$\rightarrow$old). In the early stage of evolution, the acceleration dominates the spectral evolution. At later stage of evolution, the cooling dominates over the acceleration process \citep{Tra11}. At the same time, we also pay attention to our results might have an intrinsic bias, given by the choice to fit the full low-energy bump of the SED.
 
\section{CONCLUSIONS}
We use a large sample of Fermi blazars and $\gamma$-NLS1s to study the Fermi blazar sequence and the relation between them.

 1. There is a weak anti-correlation between synchrotron peak frequency luminosity and the synchrotron peak frequency for both Fermi blazars and $\gamma$-NLS1s, which supports the blazar sequence.

 2. The Doppler-corrected peak frequency luminosity and Doppler-corrected synchrotron peak frequency is positively correlated for whole sample, which suggests that the relationship between synchrotron peak frequency and synchrotron frequency luminosity is affected by the beaming effect.

 3. There is a significant anti-correlation between jet kinetic power and the synchrotron peak frequency for both Fermi blazars and $\gamma$-NLS1s, which suggests that the $\gamma$-NLS1s could fit well into the traditional blazar sequence.

 4. There is a significant correlation between synchrotron peak frequency and synchrotron curvature for whole sample, Fermi blazars and BL Lacs, respectively. The slopes of such a correlation are consistent with statistical acceleration for the case of energy-dependent acceleration probability. For FSRQs, LBLs, IBLs, HBLs, and $\gamma$-NLS1s, we also find a significant correlation, but in these cases the slopes can not be explained by previous theoretical models.    

5. The slope of relation between synchrotron peak frequency and synchrotron curvature in $\gamma$-NLS1s is large than that of FSRQs and BL Lacs. This result may imply that the cooling dominates over the acceleration process for FSRQs and BL Lacs, while the acceleration dominates over the cooling process for $\gamma$-NLS1s. 

\acknowledgements
We are very grateful to the referee for the very helpful comments and suggestions. This work was support from the research project of Qujing Normal  University (2105098001/094).This work is supported by the National Natural Science Foundation of China (NSFC 11733001).

\end{document}